\documentstyle[12pt]{article}

\if@twoside  m
\else
\fi
\renewcommand{\theequation}{\arabic{section}.\arabic{equation}}

\newcommand{\ie}{{\em i.e.}}
\newcommand{\eg}{{\em e.g.}}
\newcommand{\cf}{{\em cf. }}
\newcommand{\etc}{{\em etc. }}
\newcommand{\QED}{\mbox{\rule[-1.5pt]{6pt}{10pt}}}
\newcommand{\lhs}{{\em lhs }}
\newcommand{\rhs}{{\em rhs }}
\newcommand{\Ran}{{\rm Ran\,}}
\newcommand{\Tr}{{\rm Tr\,}}
\newcommand{\dist}{{\rm dist\,}}
\newcommand{\im}{{\rm Im\,}}
\newcommand{\sgn}{{\rm sgn\,}}
\newcommand{\spect}{\sigma}
\newcommand{\C}{C\!\!\!\rule[.5pt]{.7pt}{6.5pt}\:\:}
\newcommand{\R}{I\!\!R}
\newcommand{\N}{I\!\!N}
\newcommand{\HH}{{\cal H}}
\newcommand{\JJ}{{\cal J}}
\newcommand{\MM}{{\cal M}}
\newcommand{\OO}{{\cal O}}
\newcommand{\RR}{{\cal R}}
\newcommand{\SS}{{\cal S}}
\newcommand{\UU}{{\cal U}}
\newcommand{\eps}{\varepsilon}
\newtheorem{claim}{Claim}[section]
\newtheorem{theorem}[claim]{Theorem}
\newtheorem{proposition}[claim]{Proposition}
\newtheorem{corollary}[claim]{Corollary}
\newtheorem{remarks}[claim]{Remarks}

\newcommand{\perturb}{{\cal{U}}_{\theta}(B,\lambda)}

\newcommand{\proof}{{\em Proof\/:}}

\begin{document}

\title{Open quantum dots: resonances from perturbed symmetry
and bound states in strong magnetic fields}

\date{}

\author{Pierre Duclos$^{a,b}\!$, Pavel Exner$^{c,d}\!$, Bernhard
Meller$^{e}$\footnote{Current address: Donaustr. 105, 12043
Berlin, Germany} }

\maketitle

\begin{quote}

{\small {\em a) Centre de Physique Th\'eorique, C.N.R.S., F-13288
Marseille Luminy,
 \\ b) PHYMAT, Universit\'{e} de Toulon et du Var,
F-83957 La Garde Cedex,}
 \\ \phantom{a) }\texttt{duclos@univ-tln.fr}\\ {\em
c) Nuclear Physics Institute, Academy of Sciences, CZ--25068 \v
Re\v z \\ \phantom{a) }near Prague,
 \\ d) Doppler Institute, Czech Technical
University, B\v rehov{\'a} 7, \\ \phantom{a) }CZ-11519 Prague,}
 \\ \phantom{a) }\texttt{ exner@ujf.cas.cz}
 \\ {\em e) Facultad de F\'\i{}sica, P.
Universidad Cat\'olica de Chile, Casilla 306, \\ \phantom{a)
}Santiago 22, Chile}
 \\ \phantom{a) }\texttt{bmeller@chopin.fis.puc.cl} }

\end{quote}

\begin{abstract}
\noindent We discuss the N\"ockel model of an open quantum dot,
\ie, a straight hard-wall channel with a potential well. If this
potential depends on the longitudinal variable only, there are
embedded eigenvalues which turn into resonances if the symmetry
is violated, either by applying a magnetic field or by
deformation of the well. For a weak symmetry breaking we derive
the perturbative expansion of these resonances. We also deduce a
sufficient condition under which the discrete spectrum of such a
system (without any symmetry) survives the presence of a strong
magnetic field. It is satisfied, in particular, if the dot
potential is purely attractive.
\end{abstract}


\section{Introduction}

A rapid progress of mesoscopic physics sheds a new light on some
traditional aspects of quantum dynamics. A particular situation
which we will consider in this paper concerns resonances which owe
their existence to a weak violation of a certain symmetry. Such
situations are common in nuclear and particle physics. In
mesoscopic systems, however, the symmetry in question refers
typically rather to geometry of the problem than to internal
quantum numbers. Hence its violation is better accessible to an
experimental investigation.

Quantum wire systems offer variety of situations with bound states
embedded into a continuum. An illustration of this claim --
aesthetically perfect -- is the second eigenvalue in a
cross--shape structure \cite{SRW, ESS}. There are many more
examples. For instance, it is easy to see using a variational
argument that {\em any} symmetric protrusion of a strip modelling
a quantum wire will possess at least one embedded eigenvalue; to
this aim one has to consider a ``half'' of it, \ie, a strip whose
one boundary is Dirichlet and deformed while the other is Neumann
and straight, and mimick the argument of \cite{BGRS}. Similar
embedded eigenstates can be produced by other interactions as well
\cite{EGST}. Moreover, the effect is not restricted to quantum
wires: in acoustic waveguides where the discrete spectrum is
absent automatically, trapped modes of a similar origin have been
studied recently \cite{DP,ELV},\footnote{When the symmetry is
violated, these trapped modes change into resonances similar to
those studied in the present paper. This effect was recently
discussed in \cite{APV}. } related states were found in elasticity
\cite{RVW}, etc.

Since the mentioned eigenvalues exist only due to a particular
symmetry, they turn into resonances once the symmetry is
violated. Naturally, there is a spectral concentration: the
resonance width measures the departure from the ideal state. A
fresh discussion of this effect with a numerical analysis and
experimental results can be found in \cite{AVD} for wires with
double stubs,\footnote{For completeness, let us recall that
double--stub wires and similar systems have also ``non-magnetic"
resonances --- see \cite{BJ,BBJ,ISY} and a recent paper in the
same vein \cite{NR}. In our model the latter are included in
eq.~(\ref{transf spec2}) below.}
 another recent example are magnetoresonances in a wire containing
a circular potential well \cite{NN}. The idea, however, is not
new. The influence of a magnetic field on the cross--structure
embedded eigenstate was treated already in \cite{SWR}. Later
N\"ockel \cite{No} investigated the case of a wire with ``quantum
dot" modeled by a potential well and the resonances which appear
in this setting when a magnetic field is applied. In distinction
to \cite{NN} he considered a rectangular well independent of the
transverse variable. In such a case there are many more resonances
because instead of the even--odd decomposition of the wire
wavefunctions now every ``longitudinal" bound state gives rise to
a series of embedded eigenvalues.

The numerical analysis used in these papers is illustrative in
revealing basic features of such resonances. At the same time, it
remains somewhat hidden in this approach that the effect exhibits
general properties and allows for a sensible perturbation theory.
In this paper we want to present a discussion which puts emphasis
on this aspect of the problem. For the sake of simplicity, we
shall do it in the N\"ockel model which is mathematically more
accessible; a treatment of the protruded wire case is left to a
subsequent publication.

Our method is based on a complex scaling the idea of which comes
from the seminal paper of Aguilar and Combes \cite{AC} (see also
\cite[Sec.~XII.6]{RS}). It has to be combined with the
transverse--mode decomposition as in \cite{DEM,DES}. However,
the present case is easier. The embedded eigenvalue turns into a
resonance again as a consequence of the tunneling between the
transverse modes, but since the latter is controlled now by an
``external" parameter such as a magnetic field or a deformation
potential, it gives rise to a power series in the appropriate
parameter which is at weak coupling dominated by the lowest term
representing the Fermi golden rule. We will describe the used
complex scaling transformation in Section~3 below. Before doing
that we shall describe the model in Section~2, list the
hypotheses and describe necessary preliminaries. As in
\cite{DES, DEM} we restrict ourselves to the situation when the
embedded eigenvalues are simple. The results can be extended to
the degenerate case without any principal difficulty, but the
resulting formulae are rather cumbersome and not very
illustrative.

In the concluding section we are going to discuss another aspect
of such a model. It concerns the opposite extremal situation: we
shall ask whether the discrete spectrum persists in strong
magnetic fields. We will derive a sufficient condition for that
which provides an affirmative answer, in particular, if the
potential describing the quantum dot is purely attractive.


\setcounter{equation}{0}

\section{Description of the model}

Let us first describe the model. The quantum wire is regarded as
an infinite planar strip. The corresponding state Hilbert space is
$\,\HH:=L^2(\Omega)\,$, where $\,\Omega:= \R\times S\,$, where
either $\,S=(-a,a)\,$, or $\,S=\R\,$ in the case when the lateral
confinement is realized by a potential alone. Given real--valued
measurable functions $\,V\,$ on $\,\R\,$, $\;W\,$ on $\,S\,$, and
$\,U\,$ on $\,\Omega\,$, we use
\begin{equation} \label{Hamiltonian}
H(B,\lambda)\,:=\, \left( -i\partial_x -By\right)^2 +V(x) -\partial_y^2
+W(y) +\lambda U(x,y)
\end{equation}
as the model Hamiltonian describing an electron in the quantum wire
subject to a homogeneous magnetic field $\,B\,$ perpendicular to the
plane. As usual we put $\,\hbar= 2m= 1\,$, the chosen sign of the
vector potential corresponds to an electron, $\,e=-1\,$, and the
field intensity $\,\vec B\,$ pointing up.

The Landau gauge employed in (\ref{Hamiltonian}) suits to our
situation in which a preferred direction exists. If $\,S=(-a,a)\,$
the transverse kinetic energy $\,-\partial_y^2\,$ operator is
supposed to satisfy the Dirichlet condition at the strip boundary,
\begin{equation} \label{Dirichlet}
\psi(x,\pm a)\,=\,0 \qquad {\rm for\;any} \qquad x\in\R\,,
\end{equation}
unless, of course, $\,W(y)\,$ tends to $\,+\infty\,$ at $\,x=\pm
a\,$ fast enough to ensure the essential self--adjointness on
$\, C^{\infty}_{0}(S)\,$; with the usual abuse of notation we
identify $\,-\partial_y^2\,$ with $\,I\otimes
(-\partial_y^2)\,$, \etc Assumptions about the potentials will
be collected below.

If $\,B=0\,$ or $\,\lambda=0\,$ we drop the corresponding symbol
at the \lhs of (\ref{Hamiltonian}), in particular,
$\,H(0):=H(0,0)\,$ is the unperturbed Hamiltonian. Since the
latter is of the form $\,h^V\otimes I +I\otimes h^W\,$, its
spectrum is the ``sum" of the corresponding component spectra.
This is a typical situation when embedded eigenvalues due to
symmetry may arise.

Our aim is to analyze how such eigenvalues turn into resonances
if the symmetry of the system is violated either by switching in
the magnetic field, or geometrically by the potential $\,U\,$
which does not decompose, $\,U(x,y)\ne U_1(x)\!+\! U_2(y)\,$ for
some $\,U_1,\, U_2\,$. We shall show the existence of
resonances, derive the corresponding perturbation series, and
compute explicitly the lowest non--real term given by the Fermi
golden rule.

Let us first list the assumptions about the potentials. Our goal
is to illustrate the mechanism of resonance production rather than
proving a most general result, and therefore we adopt relatively
restrictive hypotheses. The lateral confinement is supposed to be
``strong enough":
\begin{description}
\item{\em (i)} $\;W(y)\ge cy^2\,$ for some $\,c\ge 0\,$. In particular,
$\,c\,$ is strictly positive if $\,S=\R\,$; without loss of
generality we may assume that $\,c\ge 1\,$.
\end{description}
We make this assumption with the needs of Section~4 in mind; all
the results of Section~3 are valid under weaker requirements. For
instance, it is enough to suppose $\lim_{|x|\to\infty}
W(x)=+\infty$ which ensures that the spectrum of $\, h_{W}$,
denoted by $\,\{\nu_j\}_{j=1} ^\infty\,$, is discrete and simple,
$\,\nu_{j+1}> \nu_j\,$. Of course, it is also possible to impose
some stronger hypothesis, \eg,
\begin{description}
\item{\em (i')} $\;W(y)\ge y^2\,$ and $\,W'(y)\, \sgn y\ge c_0|y|\,$
if $\,|y|\ge y_0\,$ for some $\,c_0, y_0 > 0\,$,
\end{description}
which guarantees the existence of a uniform lower bound on the
eigenvalue spacing,
\begin{equation} \label{gap}
\inf_j \left( \nu_{j+1}\!- \nu_j\right) \,>\,0\,.
\end{equation}
The longitudinal potential $\,V\,$ which simulates the local
deformation of the quantum wire responsible for the appearance of
the discrete spectrum will be short--ranged and non--repulsive in
the mean,
\begin{description}
\item{\em (ii)} $\;V\ne 0\,$ and $\,|V(x)|\le {\rm const\,}
\langle x\rangle^{-2-\eps}\,$
for some $\,\eps>0\,$, with $\,\int_{\R} V(x)\, dx \le 0\,$,
\end{description}
where we have denoted conventionally $\,\langle x\rangle:=
\sqrt{1\!+\!x^2}\,$. Under this assumption, the longitudinal part
$\,h^V:= -\partial^2_x+V(x)\,$ of the unperturbed Hamiltonian has a
non--empty discrete spectrum,
\begin{equation} \label{long ev's}
\mu_1< \mu_2< \cdots \mu_N <0\,,
\end{equation}
which is simple and finite \cite{BGS,Kl,Ne,Si,Se}; the
corresponding normalized eigenfunctions $\,\phi_n,\;
n=1,\dots,N\,$, are exponentially decaying. It is convenient for
analyzing the resonance behaviour to adopt an analyticity
requirement. In the present case we assume that
\begin{description}
\item{\em (iii)} the potential $\,V\,$ extends to a function analytic
in a sector $\,\MM_{\alpha_0}:= \{\, z\in\C\,:\; |\arg
z|\le\alpha_0\}\,$ for some $\,\alpha_0>0\,$ and obeys there the
bound of {\em (ii).}
\end{description}
The resonances we want to study have to be distinguished from
those of the unperturbed problem coming from the operator $\,h^V$.
Fortunately this can be achieved under the last assumption: by
\cite{AC} the resonances of $h^V$ contained in $\MM_{\alpha_0}$ do
not accumulate, except possibly at the threshold. The last named
possibility does not occur if $V$ would decay exponentially, but
in fact can be ruled out with a dilation analytic potential ---
see \cite[Lemma~3.4]{J}.

Finally, the magnetic part of the perturbation is governed by a
single parameter, while the deformation is described by the
potential $\,U\,$. We shall again restrict ourselves to the
short--range case and require analyticity:
\begin{description}
\item{\em (iv)} $\;|U(x,y)|\le {\rm const\,}\langle x\rangle^{-2-\eps}\,$
for some $\,\eps>0\,$ and all $\,(x,y)\in \Omega\,$. Moreover,
$\,U(\cdot,y)\,$ extends for each fixed $\,y\in S\,$ into an
analytic function in $\,\MM_{\alpha_0}$ and satisfies there the
same bound.
\end{description}

Since $\,\sigma_{c}(h^V)=[0,\infty)\,$, the spectrum of the unperturbed
Hamiltonian consists of a continuous part,
\begin{equation} \label{sigma_c}
\sigma_c(H(0))\,=\,\sigma_{ess}(H(0))\,=\,[\nu_1,\infty)\,,
\end{equation}
and the infinite family of eigenvalues
\begin{equation} \label{sigma_p}
\sigma_p(H(0))\,=\,\left\lbrace\, \mu_n\!+\!\nu_j :\;
n=1,\dots,N\,,\; j=1,2,\dots\,\right\rbrace\,.
\end{equation}
Among these a finite subset is isolated or situated at the
threshold, while the rest satisfying the condition
\begin{equation} \label{embedding}
\mu_n+\nu_j\,>\, \nu_1
\end{equation}
is embedded in the continuous spectrum. We suppose also that
they do not coincide with the higher thresholds,
\begin{equation} \label{nothresh}
\mu_n+\nu_j\,\ne\, \nu_k
\end{equation}
for any $k$.

Finally, the transverse--mode decomposition is introduced in
analogy with \cite{DES}. The transverse eigenfunctions are denoted
as $\,\chi_j, \; h^W\chi_j=\nu_j\chi_j\,$. Then we define the
embedding operators $\,\JJ_j\,$ and their projection adjoints by
\begin{eqnarray*}
\JJ_j &\!:\!& L^2(\R) \to L^2(\Omega)\,, \qquad \JJ_j f =f\otimes
\chi_j\,, \\ \JJ_j^* &\!:\!& L^2(\Omega) \to L^2(\R)\,, \qquad
(\JJ_j^*g)(x) = \left(\chi_j, g(x,\cdot)\right)_{L^2(S)}\,.
\end{eqnarray*}
They allow us to replace the Hamiltonian $\,H(B,\lambda)\,$ by the
matrix differential operator $\,\{H_{jk}(B,\lambda)\}
_{j,k=1}^{\infty}\,$ with
\begin{eqnarray}
H_{jk}(B,\lambda) &\!:=\!& \JJ^*_j H(B,\lambda) \JJ_k \,=\,
\left( -\partial_x^2 +V(x) +\nu_j \right) \delta_{jk}
+\UU_{jk}(B,\lambda) \,, \label{matrix H} \\
\nonumber \\
\UU_{jk}(B,\lambda) &\!:=\!& 2iB\, m_{jk}^{(1)} \partial_x
+B^2 m_{jk}^{(2)} + \lambda U_{jk}(x) \,, \label{matrix U}
\end{eqnarray}
where $\,m_{jk}^{(r)}\,$ are the transverse momenta,
$$
m_{jk}^{(r)}\,:=\, \int_S y^r \overline\chi_j(y) \chi_k(y) \,dy\,,
$$
and $\,U_{jk}(x):=\int_S U(x,y) \overline\chi_j(y)
\chi_k(y) \,dy\,$.


\setcounter{equation}{0}
\section{Resonances}
\subsection{Complex scaling}

In analogy with \cite{DEM,DES} we use a complex deformation in
the longitudinal variable: we begin with the unitary operator
\begin{equation} \label{scaling}
{\cal S}_{\theta}:\; ({\cal S}_{\theta}\psi)(x,y)\,=\, e^{\theta/2}
\psi(e^{\theta}x,y)\,, \quad \theta\in\R\,,
\end{equation}
and extend this scaling transformation analytically to
$\,\MM_{\alpha_0}$. This is possible because the transformed
Hamiltonians are of the form
\begin{equation} \label{scaled H}
H_{\theta}(B,\lambda)\,:=\, {\cal S}_{\theta}H(B,\lambda)
{{\cal S}_{\theta}}^{-1}
\,=\, H_{\theta}(0)+ \UU_{\theta}(B,\lambda)\,,
\end{equation}
with $\,V_{\theta}(x):= V(e^{\theta x})\,$ and
\begin{equation} \label{scaled free}
H_{\theta}(0)\,:=\, -e^{-2\theta} \partial_x^2 -\partial_y^2
+V_{\theta}(x) +W(y)
\end{equation}
where the last named operators clearly form in view of the
assumptions {\em (ii)} and  {\em (iii)} a type (A) analytic
family of $m$--sectorial operators in the sense of \cite{Ka} for
$\,|\im\theta |<\min\{\alpha_0,\pi/4\}\,$. Furthermore
\begin{equation} \label{scaled pert}
\UU_{\theta}(B,\lambda) \,:=\, 2i\, e^{-\theta}By\, \partial_x
+B^2 y^2 + \lambda U_{\theta}(x,y)
\end{equation}
with $\,U_{\theta}(x,y):= U(e^{\theta x},y)\,$. Defining
$\,R_{\theta}(z):=(H_{\theta}(0)-z)^{-1}\,$ we prove in
Proposition~\ref{prop:A1} the following estimate
\begin{equation}
        \|{\cal U}_{\theta}(B,\lambda) R_{\theta}(\nu_{1}+\mu_{1}-1)\|\leq
        c(|B|+|B|^2+|\lambda|)\,.
    \label{eq:type A estimate}
\end{equation}
for $|\im\theta|<\min\{\alpha_0,\pi/4\}$. Thus  the ``full"
operators $\,H_{\theta}(B,\lambda)\,$ are also a type (A)
analytic family for suitably small $\,B\,$ and $\,\lambda\,$.

The transformed free part (\ref{scaled free}) separates variables, so
its spectrum is
\begin{equation} \label{transf spec1}
\sigma\left( H_{\theta}(0)\right)\,=\, \bigcup_{j=1}^{\infty}
\,\left\lbrace\, \nu_j+ \sigma\left( h^V_{\theta} \right)
\right\rbrace\,,
\end{equation}
where $\,h^V_{\theta}:= -e^{-2\theta}\partial_x^2 +V_{\theta}(x)\,$.
Since the potential is dilation analytic by assumption, the discrete
spectrum of $\,h^V_{\theta}\,$ is independent of $\,\theta\,$; we
have
\begin{equation} \label{transf spec2}
\sigma\left( h^V_{\theta}\right)\,=\, e^{-2\theta} \R_+\, \cup\,
\{\mu_1, \dots,\mu_N\} \,\cup\, \{\rho_1,\rho_2\,\dots\}\,,
\end{equation}
where $\,\rho_r\,$ are the ``intrinsic" resonances of $\,h^V\,$.
In view of {\em (iii)} and (\ref{nothresh}) the supremum of
$\,\im \rho_k\,$ over any finite region of the complex plane
which does not contain any of the points $\nu_k$ is negative, so
each eigenvalue $\,\mu_n+\nu_j\,$ has a neighbourhood containing
none of the points $\,\rho_k+\nu_{j'}\,$. Consequently, when
$\,\theta\,$ has a positive imaginary part, the eigenvalues of
$\,H_{\theta}(0)\,$ are isolated. Thus due to the relative
boundedness of $\,{\cal U}_{\theta}(B,\lambda)\,$
--- \cf(\ref{eq:type A estimate}) --- we can draw a contour
around an unperturbed eigenvalue and apply perturbation theory.
As indicated above we shall consider only the non-degenerate
case when $\,\mu_n+\nu_j\ne \mu_{n'}+\nu_{j'}\,$ for different
pairs of indices.

\subsection{Perturbation expansion}

Let us first introduce some notation. The unperturbed eigenvalue
$\mu_n+\nu_j$ will be in the following labeled as $\,e_{0}\,$.
Let $\,\theta=i\beta\,$  with an appropriately chosen
$\,\beta>0\,$; then in view of (\ref{transf spec1}) and
(\ref{transf spec2}) we may chose a contour $\,\Gamma\,$
belonging to the resolvent set of $\,H_{\theta}(0)\,$ and
encircling $\, e_{0}\,$ such that this eigenvalue is the only
one of $\,H_{\theta}(0)\,$ contained inside of $\,\Gamma\,$. As
usual, let $\,P_{\theta}\,$ denote the eigenprojection referring
to the eigenvalue $\,e_{0}\,$ and set
\[
S^{(p)}_{\theta}\,:=\,\frac{1}{2\pi i}\int_{\Gamma}^{}
\frac{R_{\theta}(z)}{(e_{0}-z)^{p}}\, dz
\]
for $\, p\geq 0 \,$. Then $\,P_{\theta}=-S^{(0)}_{\theta}\,$ and
$\,\hat R_{\theta}(z):= S^{(1)}_{\theta}\,$ is the reduced
resolvent of $\,H_{\theta}(0)\,$ at the point $\,z\,$.
We will need the following estimates:
\begin{proposition}\label{prop:PE estimate I}
    If \( \im\theta \in (0,\alpha_{0}) \),
    there exists a positive constant \( c_{\theta} \)
    such that
    \begin{description}
        \item{\em (i)}   $\;
    \max_{z\in\Gamma}\|{\cal U}_{\theta}(B,\lambda) R_{\theta}(z)\|\leq
    c_{\theta}(|B|+|B|^2+|\lambda|)\,$.
        \item{\em (ii)}
    $\;\left\|\UU_\theta(B,\lambda) S^{(p)}_{\theta}\right\|\leq
    c_{\theta}\, \frac{|\Gamma|}{2\pi}\, \left(\dist (\Gamma,e_{0})\right)^{-p}
    (|B|+|B|^2+|\lambda|) \,$ holds for $\,p\geq 0$.
    \end{description}
\end{proposition}
\proof $\;$ (i) is proved in Appendix A, the second claim
follows immediately. \quad \QED \vspace{3mm}

Now we are in position to write the perturbation expansion. By
assumption, $\, e_{0}= \mu_{n}+\nu_{j}\,$ holds for a unique
pair of the indices $\,j\,$ and $\,n\,$. Following
\cite[Sec.~II.2]{Ka} we obtain the convergent series
\begin{equation} \label{perturbed ev}
e(B,\lambda)\,=\, \mu_n+\nu_j+ \sum_{m=1}^{\infty} e_m(B,\lambda)
\,,
\end{equation}
where
\begin{equation} \label{expansion coefficients}
e_m(B,\lambda) \,=\, \sum_{p_1+\cdots+p_m=m-1} {(-1)^m \over m}\;
\Tr\, \prod_{i=1}^m\: \UU_{\theta}(B,\lambda) S^{(p_i)}_{\theta}
\end{equation}
Using the preceding proposition we infer that
\begin{equation} \label{coefficient order}
e_m(B,\lambda) \,=\, \sum_{l=0}^m\, \OO\left(B^l
\lambda^{m-l}\right)\;;
\end{equation}
in particular, $\,e_m(B)= \OO(B^m)\,$ and $\,e_m(\lambda)=
\OO(\lambda^m)\,$ in cases of the pure magnetic and pure potential
perturbation, respectively.

Let us compute first the lowest--order terms of the expansion
(\ref{perturbed ev}) in the non-degenerate case, $\,\dim
P_{\theta}=1\,$. We denote by $\,\phi_n\,$ the corresponding
eigenvector of $\,h^V\,$, \ie, $\,h^V\phi_n= \mu_n\phi_n\,$. Then
\begin{eqnarray*}
e_1^{j,n}(B,\lambda)&\!=\!& \Tr \left(\UU_{\theta}(B,\lambda)
P_{\theta} \right) \,=\, \left(\overline{\phi_n^{\theta}\otimes
\chi_j}, \UU_{\theta}(B,\lambda) \phi_n^{\theta}\otimes \chi_j
\right)
\\ \\
 &\!=\!&
\left(\phi_n\otimes \chi_j, \UU(B,\lambda) \phi_n\otimes \chi_j
\right)
\\ \\
&\!=\!& 2i B m_{jj}^{(1)} \left(\phi_n, \phi'_n \right) \,+\, B^2
m_{jj}^{(2)} +\, \lambda \left(\phi_n, U_{jj} \phi_n \right)\,.
\end{eqnarray*}
Moreover, $\,i\left(\phi_n, \phi'_n\right)= \left(\phi_n,
i\partial_x\phi_n \right)\,$ is (up to the sign) the group
velocity of the wavepacket, which is zero in a stationary state;
notice that $\,\phi_n\,$ is real-valued up to a phase factor. In
other words,
\begin{equation} \label{1st coefficient}
e_1(B,\lambda) \,=\, B^2 \int_S y^2\left| \chi_j(y)\right|^2\,
dy\,+\, \lambda \int_{\R\times S} U(x,y) \left|\phi_n(x)
\chi_j(y)\right|^2\, dx\, dy
\end{equation}
with the magnetic part independent of $\,n\,$. As usual in such
situations the first--order correction is real and does not
contribute to the resonance width. The second term in
(\ref{expansion coefficients}) can be computed in the standard way
--- see, \eg, \cite[Sec.XII.6]{RS} --- taking the limit
$\,\im\theta \to 0\,$ we get
\begin{eqnarray}\label{eq:e2}
e_2(B,\lambda) &\!=\!& -\,\Tr \left(P_{j,n} \UU(B,\lambda)
\hat{R}_{\theta\!=\!0}(e_{0}\! -i0) \, \UU(B,\lambda) P_{j,n}
\right) \\ \nonumber \\ \nonumber &\!=\!& -\,\sum_{k=1}^{\infty}\,
\left(\UU_{jk}(B,\lambda) \phi_n, \left(\left( h^V\! -\!e_{0}
+\nu_k\! -\!i0 \right)^{-1} \right) \!\hat{\phantom{|}}\:
\UU_{jk}(B,\lambda) \phi_n \right)\,.
\end{eqnarray}
We shall restrict our attention to the imaginary part of
$\,e_2(B,\lambda)\,$ which determines the resonance width to
leading order.

Notice first that the imaginary part of the last series is in fact
a finite sum. Denote $\, k_{e_{0}}:=\max\{ k:
e_{0}-\nu_{k}>0\}\,$. If the unperturbed eigenvalue is embedded
one has $\, k_{e_{0}}\geq 1\,$ by eq.~(\ref{embedding}); otherwise
the set is empty and we put $\, k_{e_{0}}=0\,$ by definition.
Introducing the symbol $\,\RR_{k}:= \left(\left( h^V\! -\!e_{0}
+\nu_k\! -\! i0 \right)^{-1} \right) \!\hat{\phantom{|}}\,$ we
have clearly $\,\RR_{k}^{\star}=\RR_{k}\,$ for $\, k>k_{e_{0}}\,$.
Consequently, the corresponding terms in the series are real and
\begin{equation}
    \im e_{2}(B,\lambda)\,=\,
    \sum_{k=1}^{k_{e_{0}}}\, \left(\UU_{jk}(B,\lambda) \phi_n,
    \,(\im\RR_{k})\, \UU_{jk}(B,\lambda)
    \phi_n \right)\,.
    \label{eq:Im e2}
\end{equation}
To make use of this formula, we have to express $\,\im\RR_{k}\,$.
We follow \cite{DES} and employ the relations
\[
\im (h^{V}-E-i0)^{-1}\,=\, \omega(E+i0)^{\star}\im
(-\partial^{2}_{x}-z)^{-1}\omega(E+i0)\,
\]
for $\, E>0\,$, where
\begin{equation}
    \omega(z)\,:=\, \Big[ I+ |V|^{1/2}(-\partial^{2}_{x}-z)^{-1}|V|^{1/2}
    \sgn(V) \Big]^{-1}\,,
    \label{eq:wave op}
\end{equation}
\ie, the inverse to the operator
\[
\Big( \omega^{-1}(z) f\Big)(x)\,:=\, f(x)
+\frac{i|V(x)|^{1/2}}{2\sqrt{z}} \int_{\R}^{}e^{i\sqrt{z}
|x-x'|}|V(x')|^{1/2}\sgn(V(x'))f(x')dx'\,.
\]
The quantity $\omega(E+i0)\,$ is well defined in view of the
assumptions {\em (ii)} and {\em (iii)}. In particular, the
latter ensures the absence of positive eigenvalues for $\,
h^{V}\,$. Furthermore,
\[
\im (-\partial^{2}_{x}-E-i0)^{-1} = \frac{\pi}{2\sqrt{E}}
\sum_{\sigma=\pm} (\tau_E^\sigma)^\star\tau_E^\sigma
\]
holds for $\, E>0\,$, where $\tau_E^\sigma:\HH^1\to\C$ is the
trace operator which acts on the first Sobolev space $\HH^1$ in
the following way,
\begin{equation}
    \tau^{\sigma}_{E} \phi := \hat\phi(\sigma \sqrt{E}),\qquad
    \sigma=\pm\,,\quad E>0,
    \label{eq: trace op}
\end{equation}
and as usual $\,\hat\phi \,$ is the Fourier transform of
$\,\phi\,$. Using the relations (\ref{eq:wave op}) and (\ref{eq:
trace op}), we can thus rewrite the expression (\ref{eq:Im e2})
as follows
\begin{eqnarray}
    \label{eq:long Im e2}
    \im e_{2}(B,\lambda) & \!=\! & \sum_{k=1}^{k_{e_{0}}}\sum_{\sigma=\pm}^{}
    \frac{\pi}{2\sqrt{e_{0}\!-\!\nu_{k}}}\Big| \tau^{\sigma}_{e_{0}\!-\!\nu_{k}}
    \omega(e_{0}\!-\!\nu_{k}\!+\! i0)\,
    {\cal U}_{jk}(B,\lambda) \phi_{n} \Big|^{2}
    \\ \nonumber \\
     & \,=\, & \sum_{k=1}^{k_{e_{0}}}\sum_{\sigma=\pm}^{}
    \frac{\pi}{\sqrt{e_{0}\!-\!\nu_{k}}}
    \left\{ -2 B^{2}|m^{(1)}_{jk}|^{2}\left|\tau^{\sigma}_{e_{0}\!-\!\nu_{k}}
    \omega(e_{0}\!-\!\nu_{k}\!+\! i0) \phi'_{n} \right|^{2}
    \right.
    \nonumber \\ \nonumber \\
&&+2\lambda B   m^{(1)}_{jk} \im
\Big(\tau^{\sigma}_{e_{0}\!-\!\nu_{k}}
    \omega(e_{0}\!-\!\nu_{k}\!+\! i0) \phi'_{n}\,,
    \tau^{\sigma}_{e_{0}\!-\!\nu_{k}}
    \omega(e_{0}\!-\!\nu_{k}\!+\! i0) U_{jk}\phi_{n}\Big)
    \nonumber \\ \nonumber \\
&&\left.
-\frac{\lambda^{2}}{2}\left|\tau^{\sigma}_{e_{0}\!-\!\nu_{k}}
    \omega(e_{0}\!-\!\nu_{k}\!+\! i0)\,U_{jk}\phi_{n}\right|^{2}
    \right\} + {\cal O}(B^{3})+ {\cal O}(B^{2}\lambda)\,,
    \nonumber
\end{eqnarray}
where we have used (\ref{matrix U}) and written explicitly only
the lowest order terms. Let us summarize the results:
 \begin{theorem}
Assume {\em (i)--(iv)} and suppose that an unperturbed
eigenvalue $\, e_{0}= \mu_{n}+\nu_{j}\,$ is simple and satisfies
(\ref{embedding}), (\ref{nothresh}). Then the perturbation with
small enough $B,\,\lambda$ changes it into a resonance; the
correponding pole position is given by (\ref{perturbed
ev})--(\ref{eq:e2}). In particular, the Fermi golden rule
(\ref{eq:long Im e2}) holds.
 \end{theorem}
 \begin{remarks}
{\rm As pointed up above the assumption {\em (i)} can be weakened.
The coefficients in (\ref{eq:long Im e2}) are generically
nonzero.}
 \end{remarks}


\setcounter{equation}{0}
\section{Bound states in a strong magnetic field}

In the final section we are going to address a different question.
Since the separation of variables and the coupling parameter
$\,\lambda\,$ are not important in the following, we merge the
potentials replacing $\,V+\lambda U\,$ by $\,U\,$ and denote
$\,H(B,\lambda)\,$ as $\,H_{U}(B)\,$.

In the absence of a magnetic field a potential well in straight
waveguide produces a non--empty discrete spectrum no matter how
shallow it is. In fact, $\,H_{\lambda U}(0)\,$ has an isolated
eigenvalue below the bottom of the essential spectrum for any small
positive $\,\lambda\,$ as long as $\,\int_{\R} U_{11}(x)
\,dx\le 0\,$, where $\,U_{11}(x):= \int_S U(x,y) |\chi_1(y)|^2 dy\,$
is the projection onto the lowest transverse mode --- the proof is
given in \cite{DE} for the hard--wall case and modifies easily to the
situation when $\,S=\R\,$ and the lateral confinement is realized by
the potential $\,W\,$.

Switching in the magnetic field changes the bound state energies;
for small $\,B\,$ the eigenvalue shift is given by (an appropriate
modification of) the perturbation theory developed above. The
eigenfunctions become complex so the bound states exhibit a
nontrivial probability current, examples are worked out in
\cite{AVD,No}. If the magnetic field is made stronger, however, it
may happen that some of the eigenvalues disappear in the
continuum; it is not apriori clear whether the increase of $\,B\,$
cannot destroy the discrete spectrum at all.

To answer this question, let us consider the operator
\begin{equation} \label{mgham}
H_U(B):=H_0(B)+U\,,\quad
H_0(B):=(-i\partial_x-By)^2-\partial_y^2+W(y)
\end{equation}
acting in $L^2(\Omega)$, $\Omega:=\R\times S$, where $S$ is either
a bounded interval $(-a,a)$ with the boundary conditions
(\ref{Dirichlet}) or $\R$. We adopt the following assumptions:
\begin{description}
\item{\em (i)} $\;W(y)\ge cy^2\,$ for some $\,c\ge 0\,$. In particular,
$\,c\,$ is strictly positive if $\,S=\R\,$; without loss of
generality we may assume that $\,c\ge 1\,$.
\item{\em (iv')} nonzero $\,U\in L^\infty(\Omega)\,$ and
$\;\lim_{|x|\to\infty}\sup_{y\in S}|U(x,y)|=0\,$.
\end{description}
Then it is easy to see that $H_U(B)$ is a well defined
self-adjoint operator and
\begin{equation} \label{infspec}
\inf\spect_{\rm ess}H_U(B)=\inf\spect_{\rm ess}H_0(B)\,.
\end{equation}
Spectral properties of the ``free'' operator $H_0(B)$ follow from
the direct integral decomposition
\begin{equation} \label{decomp}
H_0(B)=\int_\R^\oplus h_B^W(p)dp\,,\quad
h_B^W(p):=-\partial_y^2+(By-p)^2+W(y)\,,
\end{equation}
obtained by the partial Fourier transformation in the longitudinal
variable. Obviously, $h_B^W(p)$ has a simple discrete spectrum for
each $p$ and $\,\{h_B^W(p):\: p\in\R\}\,$ is a type A analytic
family. Thus we may write
\begin{equation} \label{decomp2}
h_B^W(p)=\sum_{j=1}^\infty\nu_j^B(p)\pi_j^B(p),\quad
\pi_j^B(p):=(\chi_j^B(p),\cdot)\chi_j^B(p)\,,
\end{equation}
where $\nu_j^B(p)$ and $\chi_j^B(\cdot;p)$ denote respectively the
$j$-th eigenvalue and the corresponding eigenvector of $h_B^W(p)$.
It is obvious that
\begin{equation} \label{lb1}
\nu_j^B(p)\,\ge\, \nu_j^0(0)\,=\, \nu_j\,,
\end{equation}
and moreover, an easy perturbation--theory argument shows that the
functions $\,\nu_j^B(\cdot)\,$ are continuous. If $\,S=(-a,a)\,$,
another straightforward lower bound,
\begin{equation} \label{lb2}
\nu_j^B(p)\,\ge\, \inf_{y\in S} (By\!-\!p)^2+\, \left(\pi j\over
2a \right)^2 \,,
\end{equation}
implies
\begin{equation} \label{high p}
\lim_{|p|\to\infty} \nu_j^B(p)\,=\,\infty\,.
\end{equation}
This argument does not work for $\,S=\R\,$ but (\ref{high p})
holds again, because
\begin{equation} \label{lb3}
W(y)+ (By\!-\!p)^2\,\ge\, {cp^2 \over c+B^2}
\end{equation}
holds in view of the assumption {\em (i)}. Hence for given $\,j\,$
and $\,B\,$ there is one or more values of $\,p\,$ at which the
function $\,\nu_j^B(\cdot)\,$ reaches its minimum, and
consequently
\begin{equation} \label{infess}
\inf\spect_{\rm ess}H_0(B)=\min_{p\in\R}{\nu_1^B(p)}\,.
\end{equation}
We stress that the minimum may not be unique; a simple example is
a strip with the mirror symmetry with respect to $\,y=0\,$, so
$\,\nu_j^B(p) =\nu_j^B(-p)\,$, and $\,W\,$ of a double--well form.

Our aim is to find a sufficient condition on $\,U\,$ under which
$\,H_U(B)\,$ has at least one eigenvalue below its essential
spectrum. Let $\,p_0$ be a point where the minimum of
$\,\nu_1^B\,$ is achieved. Using the following unitary equivalent
operator, $\,e^{ip_0x}H_U(B)e^{-ip_0x}=e^{ip_0x}H_0(B)e^{-ip_0x}
+U\,$, where
\begin{eqnarray*}
e^{ip_0x}H_0(B)e^{-ip_0x}&=&(-i\partial_x-By-p_0)^2-\partial_y^2+W(y)\\
&=&\int_{\R}^\oplus ( -\partial_y^2+(By-p-p_0)^2)\,dp\\
&=&\int_{\R}^\oplus \sum_{j=1}^\infty
\nu_j^B(p-p_0)\pi^B(p-p_0)\,dp\,,
\end{eqnarray*}
we find that it is enough to suppose $\,p_0=0\,$ without loss of
generality. We denote by $U^B_{1,1}(\cdot;p_0)$ the projection of
$\,U\,$ on $\,\Ran\pi^B(p_0)$\,:
\begin{equation} \label{u11}
U^B_{1,1}(\cdot;p_0):=(\chi_1^B(p_0), U(x,\cdot)
\chi_1^B(p_0))_{L^2(S)} =\int_SU(x,y)|\chi_1^B(p_0,y)|^2dy\,.
\end{equation}
Now we are in position to state a sufficient condition for
existence of the discrete spectrum.
\begin{theorem}
{\rm Assume {\em (i), (iv'),} and
\begin{equation} \label{mg attractivity}
\int_{\R} U_{1,1}(x;p_0) \,dx\,<\, 0\
\end{equation}
where $p_0$ is a minimizing value of $\chi_1^B(p_0)$; then the
discrete spectrum of $H_U(B)$ is non-empty.}
\end{theorem}
{\em Proof:} Let $\,q\,$ be the quadratic form associated with
$\,H_U(B)-\nu_1^B(0)\,$, where we suppose that the mininizing
value is reached at $p_0=0$,
\begin{equation} \label{mg form}
q[\Phi]\,:=\, \left\| (-i\partial_x -By)\Phi \right\|^2
+\,\|\partial_y \Phi\|^2 +\,(\Phi,W\Phi)\, +\,(\Phi,U\Phi)
-\nu_1^B(0)\, \|\Phi\|^2.
\end{equation}
We need only to find a trial vector $\,\Phi\in L^2(\Omega)\,$
which makes this form strictly negative. We choose it of the form
$\Phi(x,y):=\phi(x)\chi_1^B(y;0)$ with a real-valued $\phi$ from
the Schwarz space $\,\SS(\R)\,$ to be specified later. To simplify
the notations we set $\nu_1:=\nu_1^B(0)$, $\chi_1:=
\chi_1^B(\cdot;0)$ and $U_{1,1}:=U_{1,1}^B(\cdot;0)$. Then we find
\begin{eqnarray*}
\lefteqn{ \|(-i\partial_x-By)\Phi\|^2} \\ && =
\|\phi'\|_{L^2\R)}^2-
2B\,\im(\phi,\phi')_{L^2(\R)}(y\chi_1,\chi_1)_{L^2(S)}+
B^2\|\phi\|_{L^2(\R)}^2\|y\chi_1\|_{L^2(S)}^2 \\ && =
\|\phi'\|_{L^2\R)}^2+ B^2\|\phi\|_{L^2(\R)}^2
\|y\chi_1\|_{L^2(S)}^2\,,
\end{eqnarray*}
where the middle term at the \rhs vanishes for $\,\phi\,$
real--valued. We cease to indicate the scalar products involved
since it is self explanatory. Furthermore,
\begin{eqnarray*}
\|-i\partial_y\Phi\|^2+(\Phi,(W+U)\Phi)=\|\phi\|^2
\left(\|\chi'_1\|^2+(\chi_1,W\chi_1)\right)+(\phi,U_{1,1}\phi)\,.
\end{eqnarray*}
On the other hand, we have
$$ \nu_1=(\chi_1,(-\partial_y^2+B^2y^2+W)\chi_1)=
\|\chi'_1\|^2-2B(\chi_1,y\chi_1)+B^2\|y\chi_1\|^2+(\chi_1,W\chi_1)
$$
so that
$$ q[\Phi]=\|\phi'\|^2+2B(\chi_1,y\chi_1)\|\phi\|^2
+(\phi,U_{1,1}\phi)\,. $$
The second term can be ruled out as it follows from the
Feynmann-Helmann theorem:
$$ 0=\nu_1'(0)=\left.(\chi_1(p),2(By-p)\chi_1(p))\right|_{p=0}
=2B(\chi_1,y\chi_1)\,. $$
Finally we have
$$ q[\Phi]=\|\phi'\|^2+(\phi,U_{1,1}\phi)\,. $$
Now we choose a real $\,g\in\SS(\R)\,$ such that $\,g(x)=1\,$ in
$\,[-d,d]\,$ for some $\,d>0\,$ and employ the scaling trick in
analogy with \cite{GJ,DE} putting
$$ \phi_{\epsilon}(x)\,:=\, \left\lbrace \begin{array}{lll} g(x) &
\qquad \dots \qquad & |x|\le d \\ \\ g(\pm d +\epsilon(x\mp d)) &
\qquad \dots \qquad & \pm x> d
\end{array} \right.
$$
for $\eps>0$ so that
$$ q[\Phi_\eps]=\eps\|g'\|^2+(\phi,U_{1,1}\phi)\,. $$
We fix $\eps$ in such a way that
$$ \eps\|g'\|^2<{1\over2}\left|\int_{\R} U_{1,1}(x)\,dx\right| $$
and let $d$ tends to infinity, where
$$ \lim_{d\to\infty}(\phi_\eps, U_{1,1}\phi_\eps)=\int_{\R}
U_{1,1}(x)\,dx $$
holds by the dominated convergence theorem. In this way we find
vectors which make the form $\,q\,$ negative. Using the unitary
equivalence mentioned above the claim in the case $p_0\ne0$
follows easily. \quad \QED
\begin{corollary}
{\rm In addition, let $\,U\,$ be purely attractive, in other
words, nonzero with $\,U(x,y)\le 0\,$ for any
$\,(x,y)\in\Omega\,$; then $\,\sigma_{disc}(H_U(B))\,$ is
non--empty for any $\,B\,$. }
\end{corollary}
 %


\renewcommand{\thesection}{\Alph{section}}
\setcounter{section}{0}
\renewcommand{\theequation}{\Alph{section}.\arabic{equation}}
\setcounter{equation}{0}
\section{Appendix}
\begin{proposition}\label{prop:A1}
    Assume \( (i) \)--\( (iv) \).
    If \( \im\theta \in (0,\alpha_{0}) \) and $\Gamma$ is the contour
    described before Proposition~\ref{prop:PE estimate I}, there exists
     a positive constant \( c_{\theta} \) such that
    \[
    \max_{z\in\Gamma}\|{\cal U}_{\theta}(B,\lambda) R_{\theta}(z)\|\leq
    c_{\theta}(|B|+|B|^2+|\lambda|)\,.
    \]
    If \( z \) is replaced by \( z_{0}=\nu_{1}+\mu_{1}-1 \), the
    constant is independent of \( \theta \)  and the
    estimate is valid for all \( |\im\theta| <\alpha_{0} \).
\end{proposition}
\proof{}  In view of the structure of \( \perturb \) given by
(\ref{scaled pert}), we can treat the magnetic and the potential
parts, ${\cal U}_{\theta}^B$ and ${\cal U}_{\theta}^{\lambda}$, of
the perturbation separately. Furthermore, the contour \( \Gamma \)
is by assumption contained in the resolvent set of \(
H_{\theta}(0) \). Since \( R_{\theta}(\cdot) \) is bounded and
continuous and \( \Gamma \) is compact, there exists a constant \(
\tilde{c}_{\theta} \) such that
\[ \max_{z\in\Gamma}\left\|R_{\theta}(z)\right\| \leq \tilde{c}_{\theta}\,.
\]
\noindent Thus \( \max_{z\in\Gamma}\|{\cal U}_{\theta}^{\lambda}
R_{\theta}(z)\| \leq |\lambda| c_{U}\tilde{c}_{\theta}\,, \)
where \( c_{U} \) denotes a bound on the norm of \( U_{\theta}
\) which is independent of \( \theta \) by the assumption \(
(iv) \).

As for the norm \( \|{\cal U}_{\theta}^{B} R_{\theta}(z)\| \), it
is clearly sufficient to consider \( \theta=i\beta \). We have the
following estimate
\begin{equation}
    \left|{\cal U}_{i\beta}^{B}\right|^{2}=
    \left|2iBye^{-i\beta}\partial_{x}+B^{2}y^{2}\right|^{2}
    \leq 8 |B|^{2}y^{2}(-\partial_{x}^{2}) + 2 |B|^{4}y^{4}\,.
    \label{eq:magnetic perturb}
\end{equation}
in the form sense. The two terms at the \rhs will be treated
separately. Let us check first that the second one is relatively
bounded. Using a simple commutation we get the estimate
\begin{eqnarray*}
\left|y^{2}R_{\theta}(z)\right|^{2}    & \leq &
R_{\theta}(z)^{\star}\Big(-\partial_{y}^{2}+y^{2}\Big)^{2}R_{\theta}(z)
+2\left|R_{\theta}(z)\right|^{2} \\
     & \leq & \left\|h^{W}R_{\theta}(z)\right\|^{2} +2
     \tilde{c}_{\theta}^{2}\,.
\end{eqnarray*}
Since $R_{\theta}(z)$ maps $\HH$ into the domain of
$H_{\theta}(0)$ which is contained in the domain of $I\otimes
h^W$, the map $z\mapsto h^W R_{\theta}(z)$ is bounded and
continuous on the compact $\Gamma$, and therefore uniformly
bounded by some constant $c_{\Gamma,\theta}$.

To show that the first term at the \rhs of eq.~(\ref{eq:magnetic
perturb}) is relatively bounded, it remains to find a bound to \(
-\partial_{x}^{2} R_{i\beta}(z)\). One has, uniformly for \(
z\in\Gamma \),
\begin{eqnarray*}
\left\|-\partial_{x}^{2}R_{i\beta}(z)\right\|    & \!\leq\! &
\left\|-e^{-2i\beta}\partial_{x}^{2}R_{i\beta}(z)\right\|\\ &
\!\leq\! &  \|H_{i\beta}(0)R_{i\beta}(z)\|+ \|h^{W}R_{i\beta}(z)\|
+\|V_{i\beta}\|\tilde{c}_{\theta}\\
     & \!\leq\! &
     1+(e_{0}+r+\|V_{i\beta}\|)\tilde{c}_{\theta}
     +\max\{\nu_{k_{0}-1}\tilde{c}_{\theta},d_{1}\}\,,
\end{eqnarray*}
where $r$ is the radius of $\Gamma$ and we have employed the
assumption \( (iii) \) about \( V \).

It is straightforward to put these estimates together to obtain
the first claim of the proposition. For the second statement note
that \( z_{0} \) is to the left of the numerical range of \(
H_{\theta}(0) \) at the unit distance. Thus
\(\left\|R_{\theta}(z_{0})\right\| =1 \) and the constant \(
\tilde{c}_{\theta} \) in the above estimates may be replaced by 1.
Furthermore, one may bound $\|h^W R_{\theta}(z_0)\|$ independently
of $\theta$ (as long as $\im \theta<\alpha_0$) since
$$ \|h^W R_{\theta}(z_0)\| = \max_{k\in\N} \| \nu_k (h^V_{\theta}
-z + \nu_k)^{-1} \| \le \max_{k\in\N} {\nu_k\over 1+\nu_k} <
\nu_1\,. $$
 %

\subsection*{Acknowledgment}

We thank the referees for useful remarks. The research has been
partially supported by GA AS under the contract 1048801. During
the work on this paper B.M. was supported by FONDECYT, Proyecto
\#{}3970026.

\end{document}